\documentclass[twocolumn,showpacs,aps,floatfix]{revtex4}

\usepackage[latin1]{inputenc}
\usepackage{amsfonts}
\usepackage{amsmath}
\usepackage{amssymb}
\usepackage[american]{babel}
\usepackage{graphicx}
\usepackage{subfigure}
\usepackage{multirow}
\usepackage{psfrag}

\newcommand{\ra}{\rangle}
\newcommand{\la}{\langle}
\newcommand{\Z}{\mathbb{Z}}

\newcommand{\abs}[1]{\left| #1 \right|}
\newtheorem{definition}{Definition}[section]
\newcommand{\ket}[1]{|#1\rangle}
\newcommand{\bra}[1]{\langle#1|}

\newcommand{\scalar}[2]{\langle#1|#2\rangle}

\begin{document}

\title{Group Velocity of Discrete-Time Quantum Walks}
\author{A.~Kempf$^{1,2}$ and R.~Portugal$^{2,3}$}
\affiliation{$\mbox{}^1$Department of Physics, University of
Queensland, St. Lucia 4072, QLD, Australia} %
\affiliation{$\mbox{}^2$Department of Applied Mathematics,
University of
Waterloo, Waterloo, Ontario, Canada N2L 3G1} %
\affiliation{$\mbox{}^3$Laborat\'{o}rio Nacional de
Computa\c{c}\~{a}o Cient\'{\i}fica, Av.~Get\'{u}lio Vargas 333,
Petr\'{o}polis, RJ, Brazil 25651-075}

\date{\today}

\begin{abstract}
We show that certain types of quantum walks can be modeled as
waves that propagate in a medium with phase and group velocities
that are explicitly calculable.  Since the group and phase
velocities indicate how fast wave packets can propagate causally,
we propose the use of these wave velocities in a new definition
for the hitting time of quantum walks. The new definition of
hitting time has the advantage that it requires neither the
specification of a walker's initial condition nor of an arrival
probability threshold. We give full details for the case of
quantum walks on the Cayley graphs of Abelian groups. This
includes the special cases of quantum walks on the line and on
hypercubes.
\end{abstract}
 \maketitle

\section{Introduction}

While classical random walks have found important applications in
classical computing, quantum random walks are anticipated to lead to
significant applications in quantum computing. We will here focus on
discrete-time quantum walks. These were first introduced by Aharonov
\textit{et al}~\cite{YAharonov} as an example of a quantum physical
process that exhibit striking differences from classical processes.
Such differences have motivated a search for quantum algorithms
based on quantum walks that outperform their classical
counterparts~\cite{Shenvi,Amb03,Amb05}.

Aharonov \textit{et al}~\cite{DAharonov} developed the basic theory
of quantum walks on graphs with a focus on mixing-time properties.
They proved that quantum walks mix at most polynomially faster than
classical random walks. Marquezino \textit{et al}~\cite{Marquezino}
presented an analytical expression for the mixing time of
discrete-time quantum walks on the hypercube.

Particularly important in this context is the concept of hitting
time. For classical random walks, the hitting time is unambiguously
defined as the average time the walker takes to hit the final vertex
for the first time after departing from the initial vertex. The
generalization for quantum walks is not straightforward, however,
since measurements disturb the movement of the walker, see e.g.
\cite{Kempe}. One possibility is to let the walk evolve unmeasured,
i.e., unitarily, until the arrival probability at the final vertex
is above some threshold. Another possibility is to perform a partial
measurement at each step of the walk to check whether the walker has
already reached the final vertex. Both definitions of arrival times
have drawbacks. In the first case, the walker is not confirmed to
have hit the final vertex, as there is only a probability of hitting
it. In the second case, the quantum walk has been modified by the
repeated measurements so that one is actually calculating the
hitting time of a non-unitary walk that is effectively subject to a
quantum Zeno effect. Nevertheless, at least for the symmetric walk
on the hypercube~\cite{Kempe}, the two strategies for defining the
hitting time yield similar results. We note that the hitting time
can, in general, be infinite. Krovi and Brun~\cite{KB1} analyzed the
conditions for infiniteness of the hitting time for walks on Cayley
graphs based on the second definition.

Our strategy here for defining the walker's speed is to view the
quantum walker as a wave packet which evolves according to the
Schr\"odinger wave equation and then to calculate the group
velocity. In the case of the quantum walk on the line, for
example, the problem is translation invariant. Therefore, the wave
equation can be decomposed into normal modes labeled by wave
numbers, i.e., by wavelengths. More generally, one can consider,
for example, quantum walks on graphs with a general Abelian
symmetry group. There is then a corresponding generalized Fourier
transform which leads to a normal mode decomposition labeled by
generalized wave numbers.

Using the normal mode decomposition, the unitary time evolution
operator for the basic time step can then be diagonalized as a
function of the wave numbers. The logarithm of the time evolution
operator yields the Hamiltonian, and thus the energy as a function
of the wave number. This is what we are after. The energy, divided
by, or differentiated with respect to, the wave number yields the
phase and group velocities, respectively. In this sense, quantum
walks behave like waves propagating in a medium. For unitary quantum
walks the waves are neither absorbed nor amplified, i.e., the
propagation medium is passive. When the group velocity is smaller
than the phase velocity, the group velocity should then indicate the
signal velocity, i.e., the velocity with which information can
propagate.

We can now define the notion of \textit{group velocity based hitting
time} for a quantum walk on a graph as the length of the minimal
path connecting the initial and final vertices divided by the
maximal group velocity for all wave numbers, assuming, as we will
find is the case, that the group velocity remains below the phase
velocity. We will compare the group velocity based hitting time with
the standard definitions of hitting times for quantum walks on
one-dimensional lattices and on the $n$-dimensional hypercube. For
the lattice, the hitting times coincide approximately. For the
hypercube, the group-velocity hitting time is $O(n\sqrt{n})$, which
is greater than the result from the standard definitions that is
essentially $O(n)$~\cite{Kempe}. We also analyze the group velocity
of quantum walks on Cayley graphs of finite Abelian groups.

The paper is organized as follows. In Sec.~\ref{sec:Cayley} we
review the theory of quantum walks on Cayley graphs of finite
Abelian groups and present a general procedure for calculating the
group velocity. In Sec.~\ref{sec:HT} we present definitions of
quantum hitting times. In Sec.~\ref{sec:Hypercube} we compare the
group-velocity hitting time with the standard definitions of hitting
time for a quantum walk on the hypercube, and in Sec.~\ref{sec:1D}
for a quantum walk on one-dimensional lattices. In the last Section
we draw conclusions.

\section{Quantum Walks on Cayley Graphs}
\label{sec:Cayley}

A Cayley graph encodes the structure of a discrete group. Let $G$
be a finite group and $S$ a generating set. The Cayley graph
$\Gamma(G,S)$ is a directed graph such that the vertex set is
identified with $G$ and the edge set consists of pairs of the form
$(g,g  h)$ for all $g\in G$ and $h\in S$. The Cayley graph depends
in an essential way on the generating set. It is interesting to
diminish that dependence by demanding that $S$ be a symmetric set,
that is, if $h\in S$ then $h^{-1}\in S$, where $h^{-1}$ is the
inverse of $h$. In that case, the Cayley graph is an undirected
regular graph of degree $|S|$ with no loops, where $|S|$ is the
cardinality of $S$. From now on we consider only symmetric
generating sets.

Coined quantum walks can be defined on $\Gamma(G,S)$ in the
following way. Let $\mathcal{H}_S$ be the Hilbert space spanned by
states $\ket{h}$ where $h\in S$. $\mathcal{H}_S$ is the coin or
internal space. Let $\mathcal{H}_G$ be the Hilbert space spanned by
states $\ket{g}$ where $g\in G$. $\mathcal{H}_G$ is the physical
stage where the walk takes place. The evolution operator for one
step of the walk is $U=S\circ(C\otimes I)$ where
\begin{equation}
C=\sum_{h_1,h_2\in S} C_{h_1 h_2}{\left |h_1\right
\rangle}{\left\langle h_2\right|} \label{coin}
\end{equation}
is the coin operator, $I$ is the identity operator in
$\mathcal{H}_G$, and $S$ is the shift operator given by
\begin{equation}\label{eq:S}
S{\left |h\right \rangle}{\left |g\right \rangle}={\left |h\right
\rangle}{\left |g  h\right \rangle}.
\end{equation}
As we see from the last equation, if the walker is in vertex $g$ and
the result of the coin toss is $h$, then the walker moves to its
neighboring vertex $g  h$.

The analysis of the evolution is simplified in the Fourier space.
Let us suppose that $G$ is an Abelian group. In that case, $G$ is
a direct sum of cyclic groups, that is, there are integers $N_1$
to $N_n$ such that $G$ is isomorphic to
$\mathbb{Z}_{N_1}\times\dots\times\mathbb{Z}_{N_n}$, where
$\mathbb{Z}_{N}$ is the additive group modulo $N$. Any element
$g\in G$ can be written as a $n$-tuple $(g_1,\cdots,g_n)$. Such
decomposition can be determined efficiently \cite{Mosca}. The
Fourier transform on $G$ is given by the
operator %
\begin{equation}
F_G=\frac{1}{\sqrt{|G|}}\sum_{g,h\in G}\chi_g(h)\ket{g}\bra{h},
\label{eq:FT}
\end{equation} %
where $\chi_g$ is a character of $G$ given by %
\begin{equation}
\chi_g(h)=\prod_{j=1}^n\omega_{N_j}^{g_j h_j}, \label{eq:character}
\end{equation} %
where $\omega_{N_j}=\exp(\frac{2\pi i}{N_j})$ is the $N_j$-primitive
root of unity.

The Fourier basis is an orthonormal set of vectors defined by
\begin{equation}
\ket{\tilde{\chi}_h}=\frac{1}{\sqrt{|G|}}\sum_{g\in
G}\chi_g(h^{-1})\ket{g}, \label{eq:Fbasis}
\end{equation} %
where $h\in G$. That basis is interesting because any vector
$\ket{h}\ket{\tilde{\chi}_{g}}$ is an eigenvector of the shift
operator, in fact
\begin{equation}
S\ket{h}\ket{\tilde{\chi}_{g}}=\chi_h(g)\ket{h}\ket{\tilde{\chi}_{g}},
\label{eq:Sonchi}
\end{equation} %
which can be proved by using Eq.~(\ref{eq:character}). If we analyze
the form of the evolution operator, we conclude that in the Fourier
basis it acts non-trivially only on the coin subspace. So, let us
proceed with a reduced version of the evolution operator $U_g$ that
acts on states $\ket{\Psi_g(t)}=\scalar{\tilde{\chi}_g}{\Psi(t)}$,
where $\ket{\Psi(t)}$ is the generic state of the walk at time $t$
given by %
\begin{equation}
\ket{\Psi(t)}=\sum_{h\in S}\sum_{g\in
G}\tilde{\psi}_{h,g}(t)\ket{h}\ket{\tilde{\chi}_g}.
\label{eq:genstate}
\end{equation}
$U_g$ acts on states $\ket{\Psi_g(t)}=\sum_{h\in
S}\tilde{\psi}_{h,g}(t)\ket{h}$ and the matrix components are
$\bra{h_1}U_g\ket{h_2}=\chi_{h_1}(g)C_{h_1 h_2}$.

The evolution equation is given by
$\ket{\Psi_{g}(t+1)}=U_g\ket{\Psi_{g}(t)}$, which can be solved
recursively yielding $\ket{\Psi_{g}(t)}=(U_g)^t\ket{\Psi_{g}(0)}$.
One may calculate $(U_g)^t$ by diagonalizing the matrix $U_g$. Let
us call $\exp(i \omega_g^{(h)})$, $h\in S$ the eigenvalues of $U_g$.
We can calculate the Hamiltonian $H_g$ using that $U_g=\exp(-i H_g)$
after taking $\hbar=1$. In the eigenbasis of $U_g$, we have
$H_g={\rm diag}\{\omega_g^{(h)}\}$.

In Eq.~(\ref{eq:Fbasis}), parameters $h$ and $g$ play a dual role.
Parameter $g$ plays the role of the spatial position while parameter
$h$ can be considered a generalized wave number. The differentiation
of the energy $\omega_g^{(h)}$ with respect to $h$, or some quantity
directly obtained from $h$, defines the group velocity. The ratio of
the energy to $h$ defines the phase velocity.


\section{Hitting Times}
\label{sec:HT}

Using the group velocity, $v_g$, we can calculate the traveling time
from vertex $g_1$ to $g_2$. Taking edges of length one, the time is
given by $d/v_g$, where $d$ is the length of the shortest path
connecting the vertices. We define the \textit{group-velocity
hitting time} as the length of the shortest path divided by the
maximal value of the group velocity.

It is interesting to compare that ``physical'' hitting time notion
with the mathematical definition that generalizes the well known
classical hitting time notion. In the classical case, the evolution
is governed by a stochastic matrix and the hitting time is the
expected time the walker takes to hit vertex $g_2$ for the first
time starting from vertex $g_1$. In the quantum case, there is more
than one notion of quantum hitting time~\cite{Kempe,KB2}. Either one
lets the walk evolve unitarily after leaving from vertex $g_1$ and
checks when the probability at vertex $g_2$ is above some threshold,
or one performs a partial measurement at each step to measure when
the walker has reached vertex $g_2$. The first notion has the
following definition.
\begin{definition}[{\bf One-shot hitting
time}]\label{def:OSHT} Given a threshold $0\leq p \leq 1$ and a
initial condition $\ket{\phi_0}$ for the coin state, the one-shot
hitting time from vertex $g_1$ to $g_2$ of the discrete-time quantum
walk $U$ is
$$\min\{T|\,\,\ket{\phi_0}\in \mathcal{H}_S :
\sum_{h\in S}\abs{\bra{h}\la g_2|U^T|\phi_0\ra\ket{g_1}}^2 \geq p\}.$$
\end{definition}
The hitting time may be infinite if one chooses $p$ too high. On the
other hand, it is advisable to take $p$ as high as possible to have
a good chance to find the walker on vertex $g_2$.

The second notion has two definitions. Let us first define the
concurrent hitting time~\cite{Kempe}.
\begin{definition}[{\bf Concurrent hitting
time}]\label{def:CHT} A discrete-time quantum walk $U$ has a
$(T,p)$ concurrent hitting time from vertex $g_1$ to $g_2$, if the
$\ket{g_2}$-measured walk from $U$ with the initial state
$\ket{\phi_0}\ket{g_1}$ has a probability greater or equal to $p$
of stopping at a time $t\le T$.
\end{definition}
A walk is called $\ket{g}$-measured when we perform a measurement at
each step of the evolution with the projectors
$P=I\otimes\ket{g}\bra{g}$ and $Q=I-P$. If $P$ is measured the
process stops, otherwise the iteration is continued.

Krovi and Brun~\cite{KB2} proposed an alternative definition which
does not have a threshold $p$.
\begin{definition}[{\bf Average hitting
time}]\label{def:AHT} A discrete-time quantum walk $U$ with
initial state $\rho_1=\ket{\Psi}\bra{\Psi}$ where
$\ket{\Psi}=\ket{\phi_0}\ket{g_1}$ has a $(g_1,g_2)$ average hitting time %
$$\sum_{t=1}^\infty t p(t)$$ %
where %
$$p(t)={\rm Tr}\{P U (Q U)^{t-1} \rho_1 (U^\dagger Q)^{t-1} U^\dagger
P\},$$ $P=I\otimes\ket{g_2}\bra{g_2}$ and $Q=I-P$.
\end{definition}
Note that the wave function is not renormalized after the
measurement at each step.


A drawback of both the one-shot and the concurrent hitting times is
that they depend on a choice of threshold probability. Intuitively,
the threshold should not be chosen too low, because else the hitting
times would reflect the arrival of mere traces of probability.
Exponentially suppressed traces of probability often arrive quickly
but in practice cannot be considered a useful criterion for the
arrival of the walker. The threshold probability also should not be
set too high, as this could lead to an infinite hitting time. Apart
from these arguments it appears difficult, however, to further
constrain any choice of threshold probability.

Let us consider, therefore, that, intuitively, the walker carries
information and that it is necessary to wait for this information to
arrive at the final vertex. Information travels in a medium with
what is called the signal velocity which, for normal dispersion,
should be given by the maximum value of the group velocity. This is
the case here, as the analysis for walks on the line and on the
hypercube will show. The medium is passive, since the evolution is
unitary, and the dispersion relations are well behaved with the
group velocity staying below the phase velocity.

\section{Quantum Walks on Hypercubes}
\label{sec:Hypercube}

The $n$-dimensional hypercube is the Cayley graph of the group
$\Z_2^n$. Let us represent the group elements by binary $n$-tuples
$x=(x_{n-1}, \ldots x_1,x_0)$ and the generating set by \{$e_j$, $0
\le j < n$\}, where $e_j$ has a single $1$ entry in the $(n-j)^{th}$
component. In this case the vertices $(0,\cdots,0)$ and
$(1,\cdots,1)$ are opposite corners. The shift operator (\ref{eq:S})
reduces to
\begin{equation}\label{eq:Shypercube}
S{\left |e_j\right \rangle}{\left |x\right \rangle}={\left
|e_j\right \rangle}{\left |x\oplus e_j  \right \rangle},
\end{equation}
where $\oplus$ is the $n$-tuple binary sum. The character is given
by $\chi_x(e_j)=(-1)^{x_j}$ and the matrix elements of the reduced
evolution operator are $\bra{e_i}U_k\ket{e_j}=(-1)^{k_i}C_{i\,j}$.

From this point on, let us particularize the analysis to the
$n$-dimensional Grover coin, $C_{i\,j}\equiv 2/n-\delta_{i j}$,
which obeys the permutation symmetry of the hypercube. For this
coin, we can calculate explicitly the eigenvalues of $U_k$. They are
given in the following table~\cite{Marquezino}.

\begin{center}
\begin{tabular}{|c|c|}
\hline Hamming weight & Eigenvalue \\\hline
\multirow{2}{*}{$|k|=0$}  & $-1$  \\
                          & $1$  \\\hline
\multirow{4}{*}{ $1\le|k|\le n-1$} & $-1$ \\
              & $1$ \\
              & $e^{i\omega_k}$ \\
              & $e^{-i\omega_k}$ \\\hline
\multirow{2}{*}{$|k|=n$} & $1$ \\
              & $-1$\\\hline
\end{tabular}
\end{center}%
The quantity $\omega_k$ is defined by
\begin{equation}\label{omega-k}
\cos\omega_k\equiv 1-\frac{2|k|}{n}.
\end{equation}
Notice that the eigenvalues depend only on $n$ and on the Hamming
weight of $k$, defined as $|k|\equiv\sum_{j=0}^{n-1} k_j$.

One may define the velocity of a classical walker as the derivative
of the Hamming distance as function of time. We
define accordingly the group velocity of the quantum walker as%
\begin{equation}
v_{g}=\frac{d\,\omega}{d |k|},
\end{equation} %
where $\omega$ is the angular frequency. By examining the eigenvalue
table, we see that the group velocity is not zero only if $0<|k|<n$,
and it is given by
\begin{equation}\label{Vg}
v_{g}=\frac{\pm 1}{\sqrt{|k| (n-|k|)}}.
\end{equation}

\begin{figure}[h]
    \centering
    \psfrag{Vgg}{$v_{g}$}
    \psfrag{k}{\hspace{0.9cm}$|k|$}
    \includegraphics[height=7cm]{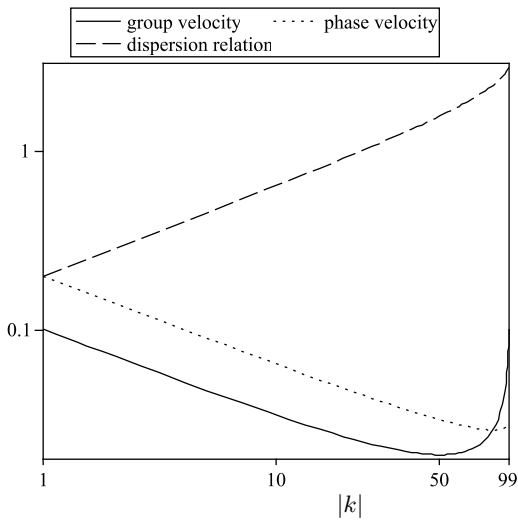}
    \caption{Group velocity, phase velocity and $\omega_k$ of the quantum
    walk on the hypercube with $n=100$ as function of $|k|$. The axes
    are in log scale.} %
    \label{fig:VgHypercube}
\end{figure}

Fig.~\ref{fig:VgHypercube} depicts $v_{g}$ as function of the wave
number when $n=100$. The maximum velocity is $1/\sqrt{n-1}$ when
$|k|=1$ or $|k|=n-1$ and the minimum is $2/n$ when $|k|=n/2$.
Fig.~\ref{fig:VgHypercube} also depicts the phase velocity and the
dispersion relation. For $k<85$ the phase velocity is greater than
the group velocity. For small $k$ the dispersion relation has
negative concavity and for $15<k<85$ it is close to a straight
line. Those facts indicate that the maximal group velocity which
is achieved at $k=1$ is the signal velocity. Then the time for the
walker to go to the opposite corner of the hypercube is
$n/v_{g}^{max} \approx n \sqrt{n}$ when $n$ is large. It is
interesting to compare this time with the current definitions of
hitting time.

Taking $\frac{1}{\sqrt{n}} \sum_{j=0}^{n-1} |e_j\ra \otimes
|0,\cdots,0\ra$ as the initial condition, the one-shot hitting time
from vertex $(0,\cdots,0)$ to $(1,\cdots,1)$ is either
$\lfloor\frac{\pi}{2}n\rfloor$ or $\lceil\frac{\pi}{2}n\rceil$ for
$p=1-O(\frac{\log^3 n}{n})$ with the condition that the hitting time
and $n$ have the same parity~\cite{Kempe}. Note that, as discussed
in \cite{Kempe}, one cannot increase the threshold probability
beyond $p=1-O(\frac{\log^3 n}{n})$ without getting infinite one-shot
hitting times, because the threshold probability is very close to
the maximum value of the probability distribution at the final
vertex.

Now when $n$ is large, with high probability the walker hits the
opposite corner at time $O(n)$. This is faster than the $O(n
\sqrt{n})$ scaling of the group-velocity hitting time. We will
discuss the interesting origin of the difference in the scaling
behavior in the last section.

Also, using the same initial condition given above, one obtains that
the walk has $(\frac{\pi}{2} n, \Omega(\frac{1}{n \log^2 n}))$
concurrent hitting time~\cite{Kempe}. Note that in this case the
probability of finding the walker at the final vertex is close to
zero for large $n$.  This result is not a contradiction with the
group-velocity hitting time for the unitary walk because the
evolution in this case is non-unitary, i.e., the walk is a different
physical process, due to the repeated measurements demanded by the
definition of the concurrent hitting time.


In comparison, also the value of the average hitting time obtained
in Ref.~\cite{KB2} is smaller than the group-velocity hitting
time.
This again is not a contradiction, because the approach of
\cite{KB2} does also not describe the same physics as we do here,
due to the non-unitary evolution caused by the repeated measurements
assumed in the definition of the average hitting time~\cite{KB2}.

\section{Quantum Walks on a 1-D Lattice}
\label{sec:1D}

In this example, the group-velocity based hitting time and the
one-shot hitting time are essentially in agreement for a suitable
choice of the threshold probability.

An one-dimensional lattice is the Cayley graph of the additive group
of integers $\mathbb{Z}$ with $S=\{1,-1\}$ as the generating set.
Since $\mathbb{Z}$ is infinite, the theory of Sec.~\ref{sec:Cayley}
does not apply straightforwardly in this case. We make the necessary
modifications in this Section.

The shift operator (\ref{eq:S}) reduces to
\begin{equation}
S{\left |j\right \rangle}{\left |n\right \rangle}={\left |j\right
\rangle}{\left |n+j\right \rangle}.
\end{equation}
Without loss of generality, we can use the Hadamard matrix as coin
operator~\cite{Nayak}, which is given by %
\begin{equation}
C=\frac{1}{\sqrt{2}}\left(
\begin{array}{rr}
1  & 1  \\ 1  & -1 \\
\end{array}
\right)\,.
\end{equation} %
The generic state of the walk is given by
\begin{equation}
{\left |\psi(t)\right \rangle}  = \sum_{j=\{1,-1\}}{
\sum_{n=-\infty}^{\infty}}\psi_{j,n}(t){\left |{j}\right
\rangle}{\left |n\right \rangle}, \label{eq:psi}
\end{equation}%
and the probability distribution by
\begin{equation}
P_{n}(t)=\sum_{j=\{1,-1\}}{\left |\psi_{j,n}(t) \right |}^2.
\label{eq:PD}
\end{equation}
The transformed amplitudes are%
\begin{equation}
\protect{\tilde{\psi}_{j,k}=\sum_{n=-\infty}^{\infty}
e^{i k n}\psi_{j,n}},
\end{equation}%
where $k\in [-\pi,\pi]$. The reduced evolution operator $U_{k}$,
which acts on
$\ket{\tilde{\psi}_{k}}=\sum_{j}\tilde{\psi}_{j,k}\ket{j}$,
is given by %
\begin{equation}
U_k=\frac{1}{\sqrt2}\left(
\begin{array}{rr}
e^{- i k}  & e^{- i k}  \\ e^{i k}  & -e^{i k} \\
\end{array}
\right)\,.
\end{equation} %
In the eigenbasis, $U_k$ is given by %
\begin{equation}
U_k=\left(
\begin{array}{rr}
\lambda_k^1  & 0 \\ 0  & \lambda_k^2 \\
\end{array}
\right)\,,
\end{equation} %
where $\lambda_k^1=e^{-i\omega_k}$ and $\lambda_k^2=e^{i(\pi
+\omega_k)}$, and $\omega_k$ is defined as the angle in
$[-\pi/2,\pi/2]$ such that $\sin(\omega_k)=\sin(k)/\sqrt{2}$.

The Hamiltonian associated with that evolution operator is%
\begin{equation} H_k=\left(
\begin{array}{rr}
\omega_k  & 0\ \ \ \  \\ 0\,  & \,\,\,\,\,\,-\pi-\omega_k \\
\end{array}
\right)\,.
\end{equation}%
We can now calculate the group velocity $v_{g}=\frac{d\,\omega}{d
k}$, where $\omega$ is the angular
frequency, obtaining %
\begin{equation}
v_{g}^\pm=\frac{\pm \cos(k)}{\sqrt{1+\cos^2(k)}}.
\end{equation} %
The phase velocity are characterized by two values, which are%
\begin{equation}
v_{ph}^+=\frac{1}{k} \arcsin(\frac{\sin k}{\sqrt{2}}) %
\end{equation}
and $v_{ph}^-=-v_{ph}^+-\pi/k$.

\begin{figure}[h]
    \centering
    \psfrag{Vgg}{$v_{g}$}
    \psfrag{k}[][]{\hspace{4cm}$k$}
    \includegraphics[height=7cm]{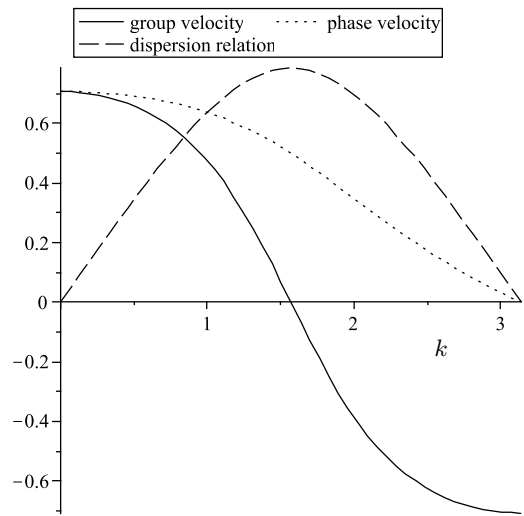}
    \caption{Group velocity, phase velocity and $\omega_k$ of the quantum walk
    on the lattice as function of $k$.} %
    \label{fig:VgLattice}
\end{figure}

Fig.~\ref{fig:VgLattice} depicts $v_{g}^+$ as function of the wave
number. The maximum velocity is $1/\sqrt{2}$ when $k=0$ and the
minimum velocity is the opposite value when $k=\pm\pi$. The phase
velocity ($v_{ph}^+$) is equal to the group velocity for $k=0$ and
is greater than the group velocity when $k>0$. The dispersion
relation has negative concavity for $0<k<\pi$. Those facts
indicate that the group velocity is in fact the signal velocity.
For the second values of phase and group velocities, we have
$v_{ph}^-<v_g^-$ when $0<k<\pi$. This anomalous case involves
velocities that are smaller or equal to the maximum group velocity
$v_g^+$ at $k=0$.

\begin{figure}[h]
    \centering
    \psfrag{V}{$V$}
    \psfrag{x}{$n$}
    \includegraphics[height=8.7cm,width=6cm,angle=270]{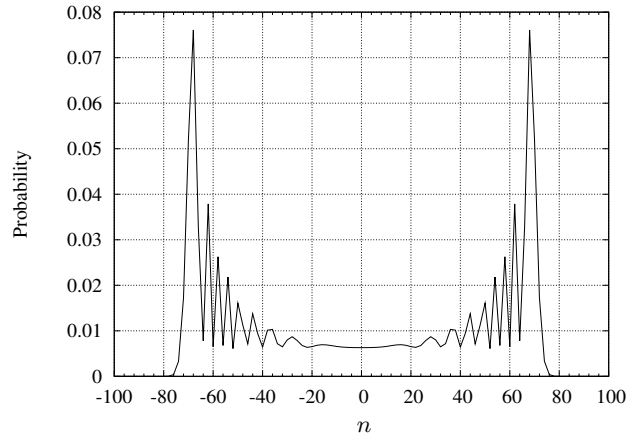}
    \caption{Probability distribution of the Hadamard walk at $t=100$ with
    initial condition $\ket{\psi(0)}=\left(\frac{\ket{0}+i \ket{1}}{\sqrt{2}}\right)\ket{n=0}$.} %
    \label{fig:Hadamard1D}
\end{figure}

It is interesting to relate the group velocity with the probability
distribution. Fig.~\ref{fig:Hadamard1D} depicts the probability
distribution of the Hadamard walk at $t=100$. Note that the
distribution is clearly non-zero in the region $-v_{g}^{max} t < n <
v_{g}^{max} t$. One can verify that the probability distribution is
not exactly zero for $|n|>v_{g}^{max} t$ although very small.

It is trivial to calculate the one-shot hitting time for the
lattice case. One can read it directly from the probability
distribution. For example, in Fig.~\ref{fig:Hadamard1D} we see
that the plot has a sharp peak at around $n_{max}=t/\sqrt{2}$. If
we take $p$ as the value $P_{n_{max}}(t)$ obtained from
Eq.~(\ref{eq:PD}), which is the most natural one to take, the
one-shot hitting time is $\sqrt{2}\,n$. In this case, the one-shot
and the group-velocity hitting time yield the same value
approximately.

The calculation of the average hitting time is somewhat tricky,
because it is defined in terms of an only slowly converging series.
Our numerical results indicate that the average hitting time is
again smaller than group-velocity hitting time.

Note that parameter $n$ for the line has a different meaning when
compared to the parameter $n$ for the hypercube. In the line, $n$ is
a linear distance to the origin while in the hypercube $n$ is a
dimension.


\section{Discussion and Conclusions}

Quantum walkers often behave similarly to wave packets in media. It
is the case, in particular, when the walker's lattice possesses an
Abelian symmetry group which allows the use of a normal mode
decomposition.

In this situation, the walker can be described as a wave packet
which over time propagates and disperses. When the walk is unitary
the walker's wave packet effectively travels in a medium which is
neither absorptive nor amplifying. For such passive media, the group
velocity is known to be a good measure of the speed with which the
wave packet can propagate information.

This motivated us to use the group velocity as the basis for a new
definition of hitting time. The new hitting time is defined as the
distance divided by the maximal group velocity for any wave number,
i.e., also for any wave packet. Therefore, among all possible
initial conditions for the quantum walker's wave packet, the
group-velocity based definition of hitting time yields the optimum.
The group velocity based hitting time also does not depend on a
choice of threshold probability. Instead, this hitting time depends
only on the coin operator and on the symmetry group which defines
the Cayley graph.

Within this approach, we calculated the hitting times for
discrete-time quantum walks on Cayley graphs of general Abelian
groups. For the special cases of the hypercubes and the
one-dimensional lattice, we compared the group velocity based
hitting times with hitting times obtained with respect to previous
definitions of the hitting time. While we found general agreement in
the case of the quantum walk on the line, we found for the
hypercubes that the group velocity based hitting times are generally
scaling slower with $n$ (the dimension) than the hitting times with
respect to previous definitions.

To explain the apparent discrepancy, let us first consider the fact
that the group velocity based hitting time is larger than the
concurrent and average hitting times. That there is a discrepancy is
not surprising, because the described physical processes are
different. In the case of the group velocity based hitting time
calculation, the quantum walk is unitary while in the other cases
the quantum walk is non-unitary due to the performance of
measurements.

More significant and interesting is the fact that, for the same
unitary quantum walk on the hypercube, the group velocity based
hitting time is larger than the one-shot hitting time.

The group velocity based hitting time is determined by how fast the
fastest wave packet can travel. In comparison, the one-shot hitting
time is based on the idea that the arrival of the walker can be
recognized by the arrival of a certain threshold probability. In the
case studied in the literature, where the walker is asked to reach
the diagonally opposite vertex in the hypercube, the threshold
probability was optimized and it is in fact close to one. It would
appear, therefore, that the arrival of such a large threshold
probability indicates the arrival of a wave packet. How, therefore,
can the group velocity based hitting time scale slower than the
one-shot hitting time?

To this end, let us consider the larger picture. In principle, in
eventual practical applications in quantum computing, it is
important how fast the walker can arrive at any vertex -- not only
the vertex opposite to the starting vertex. What, however, is the
one-shot hitting time with respect to the walker's arrival at a
vertex other than the one diagonally opposite? The calculation of
the one-shot hitting time for arrival at the diagonally opposite
vertex gives a partial answer. It was shown there that the threshold
for arrival can be chosen very high, in fact converging to $1$. The
walker is exceedingly likely to arrive there. This also shows that
the threshold for the walker's arrival at other vertices must be
chosen small in order to obtain a finite value for the one-shot
hitting time. In the larger picture, where we ask how fast a quantum
walker can visit any vertex in the graph, this indicates that the
one-shot hitting time is a difficult measure to use. This is because
suitable threshold probabilities and therefore the one-shot hitting
times can heavily depend on the end vertex considered. The reason
for that is apparently the possibility of strong destructive or
constructive interferences, that lead, in particular, to a very
significant enhancement of the arrival probability at the diagonally
opposite vertex.

The group velocity based hitting time, on the other hand, does not
require the consideration of threshold probabilities. Nor does it
seem to depend on whether or not the initial and final positions of
a wave packet that represent the walker are in a highly symmetric
relationship such as being diagonally opposed to another. In the
larger picture, where we ask how fast a quantum walker can visit any
arbitrary vertex of the graph, the group velocity based hitting time
should therefore provide a more reliable measure of that speed.

Nevertheless, for completeness, let us address the remaining
question regarding the special case of the walker arriving at the
diagonally opposite vertex. How can it be that, as the one-shot
hitting time calculation has shown, the walker can quite reliably
arrive at the diagonally opposite vertex faster than the group
velocity would indicate is possible?

To see this, let us recall that wave packets tend to possess leading
small amplitude waves that travel faster than the group velocity.
These are the Brillouin and Sommerfeld precursors, which can also be
viewed as evanescent waves. In general, in passive media, in normal
circumstances, precursors stay small during the propagation. In
active media, precursors may be amplified, thereby leading to an
apparent speed up of the wave packet. We conjecture that a similar
situation prevails here, even though the medium is passive (since
the evolution is unitary). Namely, it could be that the precursors
of the quantum walker's wave packet, on its way from the initial
vertex to the diagonally opposite vertex, constructively interfere
so as to lead to an effectively amplified precursor arriving at the
diagonally opposite vertex, before the arrival of the main wave
packet. This, and also the relationship between the group velocity
and the notion of mixing time should be interesting to explore
further. We remark that an earlier conjecture by one of us (AK) has
been confirmed, see \cite{Carteret}, that the precursor phenomenon
occurs for the quantum walk on the line.


It should also be interesting to determine the group velocities for
more general quantum walks. The method we have described here is
applicable to Cayley graphs of Abelian groups. A suitable
generalization may be applicable to non-Abelian Cayley graphs, or to
even to more general graphs that allow quantum walks, as long as
these possess some form of normal mode decomposition.
$$$$
\section*{Acknowledgments}

A.K. acknowledges the kind hospitality at the University of
Queensland and support from CFI, OIT and the Discovery and Canada
Research Chair programs of NSERC. R.P. acknowledges support from
research grant n.~2898--07--1 from CAPES.

\end{document}